\documentstyle{article}

\makeatother

\newcommand{\bee}{\begin{equation}}
\newcommand{\ene}{\end{equation}}

\newtheorem{assumption}{Assumption}[section]
\newtheorem{theorem}[assumption]{Theorem}

\def\trace{\operatorname{Tr}}
\def\operatorname#1{{\rm#1\,}}
\def\text#1{{\hbox{#1}}}
\def\BB{{\mathcal{B}}}
\def\DD{{\mathcal{D}}}
\def\EE{{\mathcal{E}}}
\def\FF{{\mathcal{F}}}
\def\GG{{\mathcal{G}}}
\def\bork{\newline\phantom{....}}
\begin{document}

\title{Asymptotics of the heat equation with `exotic' boundary conditions
or with time dependent coefficients}

\author{
   {Peter B Gilkey}\\
     {\it Mathematics Department, University of Oregon,}\\ 
              {\it Eugene Or 97403 USA}
              {\tt email:gilkey@darkwing.uoregon.edu}\\[15pt]
    {Klaus Kirsten}\\
    {\it Department of Physics and Astronomy, 
    The University of}\\
     {\it Manchester, Oxford Road, Manchester M13 9PL UK}\\
     {\tt email:~klaus@a35.ph.man.ac.uk}\\[15pt]
    {JeongHyeong Park} \\
      {\it Mathematics Department, Honam University,} \\
      {\it Kwangju 506-714 South Korea}
     {\tt email:~jhpark@honam.honam.ac.kr}\\[15pt]
    {Dmitri Vassilevich}\\
   {\it Institute for Theoretical Physics, 
University of Leipzig,}\\ {\it 04109 Leipzig, Germany.}\
    {\tt email: vassil@itp.uni-leipzig.de}
     }

\maketitle 
       
\begin{abstract} The heat trace asymptotics are discussed for operators of Laplace
type with Dirichlet, Robin, spectral, D/N, and transmittal boundary conditions.
The heat content asymptotics are discussed for operators with time
dependent coefficients and Dirichlet or Robin boundary conditions.
\end{abstract}

\def\Cal#1{{\mathcal{#1}}}
\section{Introduction}
Standard Dirichlet and Neumann boundary conditions appear in numerous
physical applications, some of which are nicely described at this
Conference. In certain cases physics requires consideration of more
`exotic' boundary value problems. For example, divergences of the
Casimir energy in non-static, but reasonably slow varying, external fields
are related to the asymptotics of the Schr\"{o}dinger equation
with a time dependent Hamiltonian. After the Wick rotation the latter
are defined by the heat trace asymptotics for operators with time
dependent coefficients. It is easy to imagine a physical experiment
when temperature of a part of the surface of a body is kept constant
while the heat flow from the other part to the outside is negligible.
Such physical experiments are described by the D/N boundary value
problem. Transmittal boundary conditions appear in the case of
semi-transparent surfaces or when the geometry of the manifold is
not smooth. The most fashionable example (and the closest
to the topic of the present Conference) of the non-smooth geometries
is given by the brane
world scenario \cite{RS}. 
Spectral boundary conditions are of relevance in one-loop quantum
cosmology and supergravity \cite{giam1,giam2}. Furthermore,
given their nice transformation properties under
chiral rotations and supertranslations there is little doubt that
study of spectral boundary conditions is also useful.

Let $D$ be an operator of Laplace type acting on the space of smooth sections to a
vector bundle $V$ over a compact Riemannian manifold $M$ of dimension $m$ with smooth
boundary $\partial M$. Let $D_\BB$ be the realization of $D$ with Dirichlet
or Robin boundary conditions; we will consider more `exotic' boundary conditions
presently. Let $e^{-tD_\BB}$ be the fundamental solution of the heat equation;
$u:=e^{-tD_\BB}\phi$ is determined by the equations:
$$u(x;0)=\phi,\ \Cal{B}u=0,\text{ and }
(\partial_t+D)u=0.$$

Let
$f$ be a smooth localizing or smearing function. We define the smeared heat trace:
$$a(f;D,\BB)(t):=\trace_{L^2}(fe^{-tD_\BB}).$$
As $t\downarrow0$ there is an asymptotic series \cite{Gr70,Gru65,S69}
\begin{equation}
\label{arefA}\quad\quad
a(f;D,\BB )\sim\textstyle\sum_{n\ge0}t^{(n-m)/2}a_n(f,D,\BB ).
\end{equation}
The asymptotic heat trace coefficients may be decomposed as the sum of an interior
and a boundary contribution:
$$a_n(f,D,\BB)=a_n^M(f,D)+a_n^{\partial M}(f,D,\BB).$$ 
The invariants $a_n^M$ and
$a_n^{\partial M}$ are computable as integrals of local geometric invariants.

Let $\psi$ be an auxiliary section to $V$ defined over the boundary $\partial M$ and
let the potential $p$ measure internal heat sources and sinks. Let $u$ be the
temperature distribution
defined by the inhomogeneous equations:
$$u(x;0)=\phi,\ \Cal{B}u=\psi,\text{ and }
(\partial_t+D)u=p.
$$
With Dirichlet boundary conditions, we keep the
boundary at constant temperature $\psi$; with Neumann boundary conditions, we pump
heat into $M$ at a rate defined by $\psi$ to control the heat flow in the normal
direction. Let
$\rho$ be the specific heat of the manifold. The total heat content
$$
\beta(p,\phi,\psi,\rho;D,\BB)(t):=\textstyle\int_Mu\rho
$$
has an asymptotic expansion as $t\downarrow0$
$$\beta\sim\textstyle\sum_{n\ge0}\beta_n(p,\phi,\psi,\rho;D,\BB)t^{n/2}.$$
The heat content asymptotics $\beta_n$ can be decomposed as the sum of an interior
and a boundary contribution given by locally computable invariants.

The coefficients $a_n$ and $\beta_n$ encode spectral information about the global
geometry of the manifold.
In Section \ref{Sect2} we
discuss the interior invariants $a_n^M$. These invariants vanish if $n$ is odd. In
Theorem
\ref{brefa}, we give formulas \cite{G75} for the invariants $a_n$ for $n=0,2,4$;
formulas for the invariants $a_6$ \cite{G75} and $a_8$ \cite{ABC89,A90} are known.

In Section \ref{Sect3}, we define the Dirichlet and Robin boundary operator - see
equation (\ref{crefaa}). In Theorem \ref{crefa}, we give formulas
\cite{BG90,KCD80,MD89} for the associated boundary correction terms
$a_n^{\partial M}$ if $n\le4$; formulas for $a_5$ are known \cite{BGKV99}. 

In Section \ref{Sect4}, we define transmittal boundary conditions - see equation
(\ref{drefaa}). In Theorem \ref{drefa}, we give formulas
\cite{BV99,GKV00,Moss00} for the boundary correction terms $a_n^\Sigma$ if
$n\le3$; formulas for $a_4$ are known \cite{GKV00}.

In Section \ref{Sect5}, we discuss spectral boundary conditions. In contrast to
Dirichlet, Robin and transmittal boundary conditions, spectral boundary conditions
are non-local. In Theorem \ref{erefa}, we give
formulas for the boundary correction terms if $n\le3$ \cite{DGK99,GK00}. Apart from
normalizing factors involving powers of
$4\pi$, the formulas of Theorems \ref{brefa}, \ref{crefa}, and
\ref{drefa} involve coefficients which are independent of the dimension $m$ of the
underlying manifold. In contrast, the formulae of Theorem \ref{erefa} are very
dimension dependent. This is one of the notable features of spectral boundary
conditions.

In Section \ref{Sect6}, we consider a time dependent family $\DD=\DD_t$ of operators
of Laplace type. The heat temperature
distribution is defined by:
$$u(x;0)=\phi,\ \Cal{B}u=0,\text{ and }
(\partial_t+\DD_t)u=0.$$
The map $\phi\rightarrow u$ is described by a smooth kernel function $K$
with the property that:
$$u(x;t)=\textstyle\int_M K(t,x,y,\DD,\BB)\phi(y)dy.$$ 
The heat trace asymptotics
are then defined not by the heat trace but directly in terms of the kernel function:
\begin{eqnarray*}
   &&a(f,\DD,\BB)(t):=\textstyle\int_Mf\trace K(t,x,x,\DD,\BB)\\
&&\qquad\sim\textstyle\sum_{n}t^{(n-m)/2}a_n(f,\DD,\BB).\end{eqnarray*}
In Theorem \ref{frefa} we give formulas for the interior invariants. 
We define boundary conditions in equation  (\ref{freffa}) which are time dependent.
In Theorem \ref{frefb}, we give formulas for the boundary correction in the heat
equation asymptotics.

In Section \ref{Sect7}, we give a brief discussion of the D/N problem
\cite{A00,D00,DGK00}. Here, in contrast to other boundary conditions, there is not a
classical asymptotic expansion at the $a_3$ level.

In Section \ref{Sect8}, we discuss the heat content asymptotics. In Theorem
\ref{hrefa}, we give formulae \cite{BDG93,BG00,G99,GP00} for the invariants $\beta_n$
for
$n\le4$ for Dirichlet or Robin boundary conditions. The coefficients which appear do
not depend on the dimension $m$. In the static setting, partial results are available
for $\beta_5$ \cite{BDG93,BG99}.

\section{Interior Heat Trace Coefficients}\label{Sect2} We introduce the following
notational conventions to describe the interior heat trace coefficients $a_n^M$. Let
Greek indices
$\mu,\nu$ range from $1$ to $m$ and index a local coordinate frame. Let Latin
indices
$i,j,k,l$ range from $1$ to $m$ and index an orthonormal frame. We adopt the Einstein
convention and sum over repeated indices. The operator $D$ determines a connection
$\nabla$ and an endomorphism
$E$ so that
$$D=-(\trace\nabla^2+E).$$
If we express
$$D=-(g^{\mu\nu}\partial_\mu\partial_\nu+a^\mu\partial_\mu+b)$$
relative to a local system of coordinates, then the connection $1$ form $\omega$ and
the endomorphism $E$ are given by:
\begin{eqnarray*}
&&\textstyle\omega_\delta=\frac12g_{\nu\delta}(a^\nu+g^{\mu\sigma} 
 \Gamma_{\mu\sigma}{}^\nu I),\text{ and}\cr
&&E=b-g^{\nu\mu}(\partial_\nu\omega_\mu 
 +\omega_\nu\omega_\mu-\omega_\sigma\Gamma_{\nu\mu}{}^\sigma).\end{eqnarray*}
If $D=\delta d$ is the scalar Laplacian, then the connection $\nabla$ is trivial and
the endomorphism $E$ vanishes. More generally, if $D=(d\delta+\delta d)_p$ is the
Laplacian on $p$ forms, then $\nabla$ is the Levi-Civita connection and $E$ is given
by the Weitzenb\"ock formulas \cite{G94}. If $D$ is the spin Laplacian, then
$\nabla$ is the spin connection 
and with our sign convention $E=-\frac14\tau$ where
$\tau$ is the scalar curvature.

Let `;' denote multiple covariant differentiation with respect to the connection
on $V$ and the Levi-Civita connection of $M$. Let
$$\rho_{ij}:=R_{ikkj}\text{ and }\tau:=\rho_{ii}$$
be the Ricci tensor and the scalar curvature. Let $\Omega$ be the curvature of the
connection $\nabla$. If ${\mathcal{A}}$ is a scalar invariant, we let
$\trace({{\mathcal{A}}}):=\trace({\mathcal{A}}I)$.

The invariants
$a_n^M$ vanish if $n$ is odd. If $n$ is even and if $n\le4$, then we have \cite{G75}:
\begin{theorem}\label{brefa}\ \begin{enumerate}
\item $a_0^M(f,D)=(4\pi)^{-m/2}\int_M\trace(f) $.
\item $a_2^M(f,D)=(4\pi)^{-m/2}\textstyle{1\over6}
     \int_Mf\trace(\tau+6E) $.
\item $a_4^M(f,D)=(4\pi)^{-m/2}\textstyle{1\over360}
    \int_Mf\trace\{60E_{;kk}+60\tau E
    +180E^2$\bork
   $+30\Omega_{ij}\Omega_{ij}+12\tau_{;kk}+5\tau^2-2|\rho|^2
     +2|R|^2\} $.
\end{enumerate}\end{theorem}

\section{Heat Trace Asymptotics for Robin and Dirichlet Boundary Conditions}
\label{Sect3}
Near the boundary, let Roman indices $a$, $b$ range from $1$ to $m-1$ and index a
local orthonormal frame $\{e_a\}$ for the tangent bundle of $\partial M$. We let
$e_m$ be the inward unit normal. Let 
$$L_{ab}:=(\nabla_{e_a}e_b,e_m)$$
be the second fundamental
form. Decompose the boundary of $M$ 
as the {\it disjoint}
union of two closed (possibly empty) sets:
$$\partial M=C_N\cup C_D.$$
Let $u_{;m}$
be the covariant derivative of $u$ with respect to the inward unit normal using the
natural connection defined by $D$. Let the boundary operator
\begin{equation}\label{crefaa}\qquad\BB u:=u|_{C_D}\oplus (u_{;m}+Su)|_{C_N}
\end{equation}
define Dirichlet boundary conditions on $C_D$ and Robin boundary conditions on $C_N$.
Let `:' denote multiple covariant differentiation of tensors defined on
$\partial M$ with respect to the connection on $V$ and the Levi-Civita
connection of the boundary. Note that  `;' and `:' differ by the second
fundamental form.  We have \cite{BG90,KCD80,MD89}:
\begin{theorem}\label{crefa}\ \begin{enumerate}
\smallskip\item $a_0^{\partial M}(f,D,\Cal{B})=0$.
\smallskip\item $a_1^{\partial M}(f,D,\Cal{B})=-(4\pi)^{(1-m)/2}\frac14
     \int_{C_D}\trace(f) +
   (4\pi)^{(1-m)/2}\frac14\int_{C_N}\trace(f) $.
\smallskip\item $a_2^{\partial M}(f,D,\Cal{B})=(4\pi)^{-m/2}\frac16\int_{C_D}
\trace\{2fL_{aa}
-3f_{;m}\}$\bork$+(4\pi)^{-m/2}\frac16\int_{C_N}\trace\{2fL_{aa}
+12fS+3f_{;m}\} $.
\smallskip\item $a_3^{\partial M}(f,D,\Cal{B})=-(4\pi)^{(1-m)/2}\frac1{384}
\int_{C_D}
\trace\{f(96E+16\tau 
-8\rho_{mm}+7L_{aa}L_{bb}$
\bork$-10L_{ab}L_{ab})-30f_{;m}L_{aa}+24f_{;mm}\}
+(4\pi)^{(1-m)/2}\frac1{384}\int_{C_N}
\trace\{f(96E+16\tau$
\bork$ -8\rho_{mm}+13L_{aa}L_{bb}
+2L_{ab}L_{ab}+96SL_{aa}
+192S^2)
+f_{;m}(6L_{aa}+96S)$\bork$+24f_{;mm}\} $.
\smallskip\item $a_4^{\partial M}(f,D,\Cal{B})=(4\pi)^{-m/2}\frac1{360}
\int_{C_D}
\trace\{f(-120E_{;m}+120EL_{aa}-18\tau_{;m}$\bork$+20\tau L_{aa}
-4\rho_{mm}L_{bb}
-12R_{ambm}L_{ab}+4R_{abcb}L_{ac}
+24L_{aa:bb}+\frac{40}{21}L_{aa}L_{bb}L_{cc}$\bork$
-\frac{88}7L_{ab}L_{ab}L_{cc}
+\frac{320}{21}L_{ab}L_{bc}L_{ac})+f_{;m}(-180E
-30\tau 
-\frac{180}{7}L_{aa}L_{bb}$\bork$+\frac{60}7L_{ab}L_{ab})+
24f_{;mm}L_{aa}
-30f_{;iim}\} 
+(4\pi)^{-m/2}\frac1{360}\int_{C_N}\trace\{f(240E_{;m}
$\bork$+120EL_{aa}+42\tau_{;m}
+24L_{aa:bb}
+20 \tau L_{aa}-4\rho_{mm}L_{bb}
-12R_{ambm}L_{ab}$\bork$+4R_{abcb}L_{ac}+\frac{40}3L_{aa}L_{bb}L_{cc}
+8L_{ab}L_{ab}L_{cc}
+\frac{32}3L_{ab}L_{bc}L_{ac}+720SE+120S \tau $\bork$
+144SL_{aa}L_{bb}
+48SL_{ab}L_{ab}+480S^2L_{aa}+480S^3+120S_{:aa})
+f_{;m}(180E$\bork$+72SL_{aa}+240S^2
+30\tau +12L_{aa}L_{bb}+12L_{ab}L_{ab}
)
+f_{;mm}(120S+24L_{aa})$\bork$+30f_{;iim}\} $.
\end{enumerate}\end{theorem}

\section{Transmittal boundary conditions}\label{Sect4}
Let $\partial M$ be empty. We suppose given
a hypersurface $\Sigma$ which divides $M$ into two smooth components $M^\pm$. We
also suppose given operators of Laplace type $D^\pm$ on $M^\pm$. Let $\nu$ be the
inward normal of $\Sigma\subset M^+$. For
$\phi=(\phi^+,\phi^-)$, we define:
\begin{eqnarray}&&\BB\phi:=\{\phi^+|_\Sigma-\phi^-|_\Sigma\}\label{drefaa}\\
   &&\quad\oplus\{(\nabla_\nu^+\phi^+)|_\Sigma
 -(\nabla_\nu^-\phi^-)|_\Sigma-\Xi\phi^+|_\Sigma\}.\nonumber\end{eqnarray}
Thus $\phi$ satisfies {\it transmittal} boundary conditions if $\phi$ is continuous
and if the normal derivatives of $\phi^+$ match to the normal derivatives of
$\phi^-$ modulo the impedance transmission term $\Xi$. We let $D_\BB$ be the
realization of $D=(D^+,D^-)$ with these boundary conditions. Let
$f=(f^+,f^-)$ be smooth on $M^\pm$ and continuous on $\Sigma$; we impose no matching
condition on the normal derivatives of $f$. Let
\begin{eqnarray*}
&&a(f,D,\BB)=\trace_{L^2}(fe^{-tD_\BB})\\
&&\quad\sim\textstyle\sum_{n\ge0}t^{(n-m)/2}a_n(f,D,\BB).
\end{eqnarray*}
We can decompose the invariants $a_n$ in the form:
\begin{eqnarray*}
&&a_n(f,D,\BB)=a_n^{M^+}(f,D)+a_n^{M^-}(f,D)\\&&\qquad+a_n^\Sigma(f,D,\BB).
\end{eqnarray*}
The invariants $a_n^{M^\pm}$ can be computed using the formulas of Theorem
\ref{brefa}. Let $\nu^\pm$ be the
inward normals of $\Sigma\subset M^\pm$; $\nu=\nu^+=-\nu^-$. We let 
$$\omega_a:=\nabla_a^+-\nabla_a^-\text{ and }
L^\pm_{ab}:=\pm(\nabla_{e_a}^\pm e_b,\nu).$$
The tensor $\omega_a$ measures the failure of the connections $\nabla^\pm$ to agree
on $\Sigma$; it is a chiral tensor - if we interchange the roles of $+$ and $-$,
then this tensor changes sign. The tensors $L^\pm$ are the second fundamental
forms of the inclusions
$\Sigma\subset M^\pm$. We refer to
\cite{GKV00} for the proof of the following theorem; see also related work in
\cite{BV99,Moss00}.

\begin{theorem}\ \label{drefa}\begin{enumerate}\item $a_0^\Sigma(f,D,\Xi)=0$.
\item $a_1^\Sigma(f,D,\Xi)=0$.
\item $a_2^\Sigma(f,D,\Xi)=
     \textstyle(4\pi)^{-m/2}\frac16\textstyle\int_\Sigma
       \trace\{2f(L_{aa}^++L_{aa}^-)-6f\Xi\}$.
\item $a_3^\Sigma(f,D,\Xi)=\textstyle(4\pi)^{(1-m)/2}\frac1{384}
\textstyle\int_\Sigma
\trace\{\frac32f(L_{aa}^+L_{bb}^++L_{aa}^-L_{bb}^-+2L_{aa}^+L_{bb}^-)
$\bork$
    +3f(L_{ab}^+L_{ab}^++L_{ab}^-L_{ab}^-+2L_{ab}^+L_{ab}^-)
    +9(L_{aa}^++L_{aa}^-)(f_{;\nu^+}^++f_{;\nu^-}^-)
$\bork$
+48f\Xi^2+24f\omega_a\omega_a
    -24f(L_{aa}^++L_{aa}^-)\Xi
    -24(f_{;\nu^+}^++f_{;\nu^-}^-)\Xi\}$.
\end{enumerate}
\end{theorem}

\section{Spectral boundary conditions}\label{Sect5}
Let $P:C^\infty(E_1)\rightarrow C^\infty(E_2)$ be an elliptic complex of
Dirac type; this means that the associated second order operators $P^*P$ and $PP^*$
are of Laplace type. Since such an elliptic complex does not in general admit local
boundary conditions \cite{APS75}, we impose {\it spectral boundary
conditions}. Let $\gamma$ be the leading symbol of the operator $P$. Then
$\gamma+\gamma^*$ defines a unitary Clifford module structure on $E_1\oplus E_2$.
Let $\nabla=\nabla_1\oplus\nabla_2$ be a compatible unitary connection \cite{BG92}.
This means that:
\begin{eqnarray*}
&&\nabla(\gamma+\gamma^*)=0,\text{ and}\\
&&(\nabla s,\tilde s)+(s,\nabla\tilde s)=d(s,\tilde s).\end{eqnarray*}
In general this auxiliary connection will not coincide with the
connections associated to the Laplacians
$\Delta_1=P^*P$ and $\Delta_2=PP^*$. 
We expand 
$$P=\gamma^i\nabla_i+\psi$$
where $\psi$ is a smooth linear map from $E_1$ to $E_2$.
We parallel translate frames for $E$ along the normal geodesic rays defined by the
inward unit normal. Relative to such a gauge, we have $\nabla_m=\partial_m$. We set
$x^m=0$ to define the {\it tangential operator} $B$ on $C^\infty(E_1|_{\partial M})$:
$$B:=(\gamma^m)^{-1}\{\gamma^\alpha\nabla_\alpha+\psi\}.$$
Let $B^*$ be the adjoint of $B$ relative to the structures on the boundary and let
$\Theta$ be an auxiliary self-adjoint operator. We define
$$A:=\frac{B+B^*}2+\Theta$$
Let $\BB$ denote projection on the span of the eigenspaces corresponding to the
non-negative spectrum of $A$. Let $P_\BB$ be the associated realization of $P$ and
let $D_\BB:=(P_\BB)^*P_\BB$. Results of Grubb and Seeley
\cite{GS93} show that there is an asymptotic series as $t\downarrow0$ of the
form:
\begin{eqnarray*}
&&\trace_{L^2}\{fe^{-tD_{{\mathcal B}}}\}\\
&\sim&
 \textstyle\sum_{0\le k\le m-1}a_k(F,D,{\mathcal B})t^{(k-m)/2}\end{eqnarray*}
modulo terms which are $O(t^{-\frac18})$.
(There is in fact a complete asymptotic series with log terms, 
but we shall only be interested in
the first few terms in the series). We shall assume $m\ge4$ so the terms $a_n$ for
$n\le3$ are well defined. Let
\begin{eqnarray*}
&&\hat\psi:=\gamma_m^{-1}\psi,\text{ and}\\
  &&C(m):=\Gamma(\textstyle\frac m2)\Gamma(\textstyle\frac12)^{-1}
\Gamma(\frac{m+1}2)^{-1}.\end{eqnarray*}
We refer to \cite{DGK99,GK00} for the proof of the following theorem:
\begin{theorem}\label{erefa} We have\begin{enumerate}
\item $a_0(f,D,{\mathcal B}) =  (4\pi )^{-m/2} \int_{M}\trace\{f\}$.
\item 
      $a_1(f,D,{\mathcal B}) =
(4\pi)^{-(m-1)/2}\frac14(C(m)-1)\int_{\partial M}\trace\{f\}$.
\item $a_2(f,D,{\mathcal B})
     =a_2^M(f,D)
   +(4\pi)^{-m/2}\int_{\partial M}f\trace\{
     \frac{1}{2}[\hat\psi+\hat\psi^*]$\bork$+\frac13(1-
     \frac34\pi C(m))L_{aa}\}
     -\frac{m-1}{2(m-2)}(1-
        \frac12 \pi C(m))\trace\{f_{;m}\}.$
\smallskip\item $a_3(f,D,{\mathcal B})
=(4\pi)^{-(m-1)/2}\textstyle\int_{\partial
M}f\trace\{
       \frac 1 {32} (1-\frac{C(m)} {m-2} ) (\hat\psi \hat\psi+\hat\psi^*
\hat\psi^* )$\bork
  $     +\frac 1 {16}  ( 5-2m+\frac{7-8m+2m^2}{m-2} C (m))\hat\psi\hat\psi^*
   +\frac 1 {32 (m-1)}(2m-3 -\frac{2m^2-6m+5}{m-2}C (m) )$\bork$
\quad\cdot(\gamma_a^T\hat\psi\gamma_a^T\hat\psi
       +\gamma_a^T\hat\psi^*\gamma_a^T\hat\psi^*)
      +\frac 1 {16 (m-1)}(1+\frac{3-2m}{m-2} C (m))
       \gamma_a^T\hat\psi\gamma_a^T\hat\psi^*$\bork$
    -\frac1{48}(\frac{m-1}{m-2}C(m)-1)\tau
    +\frac1{48}(1- \frac{4m-10}{m-2}C(m))\rho_{mm}$\bork
   $+\frac {L_{ab}L_{ab}}{48 (m+1)}(\frac{17+5m} 4 +\frac{23-2m-4m^2}{m-2} C(m)
   )
   +\frac {L_{aa} L_{bb}}{48 (m^2-1)}(-\frac{17+7m^2}{8}$\bork$
   \qquad+
\frac{4m^3-11m^2+5m-1}
         {m-2} C (m))
+\frac 1 {8(m-2)} C (m)(\Theta \Theta +\frac 1 {m-1}
     \gamma_a^T\Theta \gamma_a^T\Theta)\}$\bork
    $+\frac {L_{aa}f_{;m}}{8(m-3)} ( \frac{5m-7} 8 -\frac{5m-9} 3 
           C (m)) \trace\{I\}
    +\frac{m-1}{16(m-3)}(2C(m)-1)f_{;mm}\trace\{I\}$.
\end{enumerate}\end{theorem}

\section{Time dependent coefficients}\label{Sect6} Previously, we have considered
static operators. Let $\DD$ be an operator of Laplace type where the coefficients
are time dependent. 
We expand
$$\DD u:=Du+\textstyle\sum_{r>0}t^r\{\GG_{r,}{}_{ij}u_{;ij}
        +\FF_{r,}{}_iu_{;i}+\EE_ru\}$$
and consider time dependent Dirichlet and Robin boundary conditions:
\begin{equation}
   \BB u:=u|_{C_D}\oplus
   (u_{;m}+Su+t(T_au_{;a}+S_1u))|_{C_N}.\label{freffa}\end{equation}
We consider the heat
equation:
$$u(x;0)=\phi,\ 
  \BB u=0,\ 
  (\partial_t+\DD)u=0.$$
There is a smooth kernel function $K(t,x,y,\DD,\BB)$ so that we may express:
$$u(x;t)=\textstyle\int_MK(t,x,y,\DD,\BB)\phi(y).$$
We take the fiber trace to define
\begin{eqnarray*}
a(f,\DD,\BB)&=&\textstyle\int_Mf\trace_VK(t,x,x,\DD,\BB)\\
 &\sim&\textstyle\sum_{n\ge0}t^{(n-m)/2}a_n(f,\DD,\BB).\end{eqnarray*}
This agrees with the previous definition if $\DD$ is static.
We refer to \cite{GKP00} for the proof of the following two results which
give the additional terms in the asymptotic expansion arising from the time
dependent nature of the coefficients:
\begin{theorem}\label{frefa}\ 
\begin{enumerate}
\smallskip\item $a_0^M(f,\DD)=a_0^M(f,D)$.
\smallskip\item $a_2^M(f,\DD )=a_2^M(f,D)
+(4\pi)^{-m/2}\frac16
\int_Mf\trace(\frac32\GG_{1,ii}) $.
\smallskip\item $a_4^M(f,\DD )=a_4^M(f,D)
+(4\pi)^{-m/2}\frac1{360}\int_Mf\trace(
\frac{45}4\GG_{1,ii}\GG_{1,jj}+\frac{45}2\GG_{1,ij}\GG_{1,ij}
$\bork$+60\GG_{2,ii}-180\EE_{1}
+15\GG_{1,ii}R_{jkkj}
  -30\GG_{1,ij}R_{ikkj}$\bork$+90\GG_{1,ii}E
  +60\FF_{1,i;i}+
 15\GG_{1,ii;jj}-30\GG_{1,ij;ij}) $.\end{enumerate}
\end{theorem}

Let $\BB_0$ denote the associated static boundary conditions. We have:

\begin{theorem}\ \label{frefb}
\begin{enumerate}
\smallskip\item $a_n^{\partial M}(f,\DD,\Cal{B})=a_n^{\partial M}
(f,D,\Cal{B}_0)$ for $n\le2$.
\smallskip\item $a_3^{\partial M}(f,\DD,\Cal{B})=
\textstyle a_3^{\partial
M}(f,D,\Cal{B}_0)+(4\pi)^{(1-m)/2}\frac1{384}\int_{C_D}
    f\trace(-24\GG_{1,aa}) $
\bork$+(4\pi)^{(1-m)/2}\frac1{384}\textstyle\int_{C_N}f\trace(24\GG_{1,aa})
    $.
\par\item 
$a_4^{\partial M}(f,\DD,\Cal{B})=
\textstyle a_4^{\partial M}(f,D,\Cal{B}_0)+(4\pi)^{-m/2}
\frac1{360}\int_{C_D}f\trace\{30\GG_{1,aa}L_{bb}$\bork$ 
-60\GG_{1,mm}L_{bb}+
  30\GG_{1,ab}L_{ab}+
30\GG_{1,mm;m}-30\GG_{1,aa;m}\textstyle+0\GG_{1,am;a}-30\FF_{1,m}\}
$\bork
$+f_{;m}\trace\{-45\GG_{1,aa}+45\GG_{1,mm}\} 
\textstyle+(4\pi)^{-m/2}\frac1{360}\int_{C_N}f\trace\{30\GG_{1,aa}L_{bb}
 $\bork$+120\GG_{1,mm}L_{bb}
 -150\GG_{1,ab}L_{ab}-60\GG_{1,mm;m}
+60\GG_{1,aa;m}+150\FF_{1,m}$\bork$
+180S\GG_{1,aa} -180S\GG_{1,mm}+360S_1\}
+f_{;m}\trace\{45\GG_{1,aa}-45\GG_{1,mm}\} $.
\end{enumerate}
\end{theorem}

\section{The D/N Problem}\label{Sect7} In Section \ref{Sect3}, we assumed that
$C_N\cap C_D$ was empty to define the boundary operator $\BB$ of equation
(\ref{crefaa}). This meant that the Neumann and Dirichlet components did not
interact. In this section, we suppose that
$\Sigma:=C_D\cap C_N$ is a non--empty smooth submanifold of $\partial M$ of
dimension $m-2$. 

We can motivate this more generalized setting with a physical example. Let $M$ be
a solid ball which floats in ice water. The part of the boundary of the ball
which is in air satisfies Neumann conditions and the part which is in the water
satisfies Dirichlet conditions. Thus $\mathcal{B}$
is defined by complementary spherical caps about the north and south poles
of the ball which intersect in a circle of latitude.

The setting where $\Sigma$ is not empty is known in the literature as the
$N/D$ problem. It has been investigated extensively from the
functional analytic point of view \cite{LM72,P63,Pr81,S88}. It is
natural to conjecture the asymptotic expansion described in (\ref{arefA}) could be
generalized to this setting by adding an extra integral over
$\Sigma$ of some suitably chosen local invariant. Preliminary computations
\cite{A00,D00} suggest the additional correction term for $n=2$ is given by:
$$a_2^\Sigma=-\textstyle\frac{\pi}4(4\pi)^{-m/2}\textstyle\int_\Sigma\trace(f).$$
However, it has been shown \cite{DGK00} that the
asymptotic expansion does not exist with locally computable coefficients at the
$a_3$ level. Thus probably either log terms arise or non-local terms arise; it is
also possible, of course, that no asymptotic expansion exists.

\section{Heat Content Asymptotics}\label{Sect8} Let $\DD$ be a time dependent family
of operators of Laplace type. Let
$\psi(y;t)$ be a smooth section to
$V$ defined over $\partial M$. On the Neumann boundary component
$C_N$, we use a Neumann heat pump to pump heat into
$M$ at a rate defined by $\psi$; in this setting, the parameter $S$ controls the
coupling between the heat transfer and the temperature difference on the Neumann
component. On the Dirichlet component we use a Dirichlet heat pump to keep the
temperature at $\psi$. Let $p$ be a heat source. The temperature distribution
$u=u_{p,\phi,\psi}(x;t)$ which is defined by these data is the solution to the
equations:
\begin{eqnarray*}
&&(\partial_t+\DD)u=p,\ u(x;0)=\phi,\text{ and}\\
&&\BB u=\psi.
\end{eqnarray*}
Let $\rho$ be the specific heat; we regard $\rho$ as a section to the dual
bundle $V^*$ and let
$\langle\cdot,\cdot\rangle$ denote the dual pairing between $V$ and $V^*$. The total
heat energy content $\beta$ is defined by
$\beta(t):=\textstyle\int_Mu\rho$.
We expand $\beta$ in an asyptotic series as $t\downarrow0$ to define the
associated heat content asymptotics:
$$\beta\sim\textstyle\sum_nt^{n/2}\beta_n(p,\phi,\psi,\rho;D,\BB).$$
Let $\tilde D$ and $\tilde\BB$ be the dual operator and dual boundary
condition on the dual bundle $V^*$. We summarize results of
\cite{BDG93,BG00,G99,GP00}:
\begin{theorem}\label{hrefa}\ \begin{enumerate}
\item $\beta_0(p,\phi,\psi,\rho;\DD,\BB)=
{\textstyle\int}_{M}\langle \phi,\rho\rangle $.
\item $\beta_1(p,\phi,\psi,\rho;\DD,\BB)=
 -{2\over\sqrt{\pi}}{\textstyle\int}_{C_D}\{\langle \phi-\psi_0,\rho\rangle\} $.
\item $\beta_2(p,\phi,\psi,\rho;\DD,\BB)=
 -{\textstyle\int}_{M}\{\langle D\phi,\rho\rangle-\langle p_0,\rho\rangle\}
$\bork
 $ +{\textstyle\int}_{C_D}\{
    \langle {\frac12}L_{aa}(\phi-\psi_0),\rho\rangle-\langle
(\phi-\psi_0),\rho_{;m}\rangle\} $\bork
$ +{\textstyle\int}_{C_N}\{\langle{(\Cal B}\phi-\psi_0),\rho\rangle\} $.
\item $\beta_3(p,\phi,\psi,\rho;\DD,\BB)=
 -{2\over\sqrt{\pi}}{\textstyle\int}_{C_D}\{\frac23\langle p_0,\rho\rangle
     -\frac23\langle D\phi,\rho\rangle$\bork$
 -\frac23\langle (\phi-\psi_0),\tilde D\rho\rangle
 +{1\over3}\langle (\phi-\psi_0)_{:a},\rho_{:a}\rangle
     -\frac23\langle\psi_1,\rho\rangle
 +\langle(-\frac13E+{\textstyle{1\over{12}}} L_{aa}L_{bb}$\bork$
 -{\textstyle{1\over6}} L_{ab}L_{ab}+{\textstyle{1\over6}} R_{am am}
   -\Cal{G}_{1,mm})\cdot(\phi-\psi_0),\rho\rangle\} $\bork$
+\frac4{3\sqrt{\pi}}{\textstyle\int}_{C_N}
     \{\langle(\BB \phi-\psi_0),\tilde\BB \rho\rangle
    \} $.
\item $\beta_4(p,\phi,\psi,\rho;\DD,\BB)=
{\frac12} {\textstyle\int}_{M}\{\langle p_1,\rho\rangle-\langle Dp_0,\rho\rangle
   +\langle D\phi,\tilde D\rho\rangle$\bork
    $-\langle(\Cal{G}_{1,ij}\phi_{;ij}
     +\Cal{F}_{1,i}\phi_{;i}
      +\Cal{E}_1\phi),\rho\rangle\} 
 +{\textstyle\int}_{C_D}\{
   \frac14L_{aa}\langle p_0,\rho\rangle
    -\frac12\langle p_0,\rho_{;m}\rangle$\bork$
- {\frac14}L_{aa}\langle\psi_1,\rho\rangle 
+\frac12\langle \psi_1,\rho_{;m}\rangle
+{\frac12}\langle
 (D\phi)_{;m},\rho\rangle 
 +{\frac12}\langle(\phi-\psi_0),(\tilde D\rho)_{;m}\rangle
$\bork$  -{\textstyle{1\over4}}\langle
 L_{aa}D\phi,\rho\rangle 
 -{\textstyle{1\over4}}\langle L_{aa}(\phi-\psi_0),\tilde D\rho\rangle 
 +\langle ({\textstyle{1\over{8}}} E_{;m}- {\textstyle{1\over{16}}} L_{
ab}L_{ab}L_{cc}$\bork$+{\textstyle{1\over{8}}} L_{ab}L_{ac}L_{bc}
 -{\textstyle{1\over{16}}} R_{am bm}L_{
ab}+{\textstyle{1\over{16}}} R_{ab cb}L_{
ac}+\frac1{32}\tau_{;m}$\bork$
+{\textstyle{1\over{16}}} L_{
ab:ab})(\phi-\psi_0),\rho\rangle -{\textstyle{1\over4}} L_{
ab}\langle (\phi-\psi_0)_{:a},\rho_{:b}\rangle$\bork$ 
 -{\textstyle{1\over{8}}}\langle\Omega_{
am}(\phi-\psi_0)_{:a},\rho\rangle
 +{\textstyle{1\over{8}}}\langle\Omega
_{am}(\phi-\psi_0),\rho_{:a}\rangle$\bork
$+(\textstyle{7\over16}\Cal{G}_{1,mm;m}
      -{1\over4}\Cal{G}_{1,mm}L_{aa}
-\textstyle{5\over16}\Cal{F}_{1,m})\langle(\phi-\psi_0),\rho\rangle$
\bork
 $-\textstyle{5\over16}\Cal{G}_{1,am}\langle(\phi-\psi_0)_{:a},\rho\rangle
    +{1\over2}\Cal{G}_{1,mm}\langle(\phi-\psi_0),\rho_{;m}\rangle\}$
\bork
$   +{\textstyle\int} _{C_N}\{\frac12\langle\Cal{B}p_0,\rho\rangle
-{\frac12}\langle(\BB \phi-\psi_0),\tilde
D\rho\rangle-{\frac12}\langle D\phi,\tilde \BB \rho\rangle
-{1\over2}\langle\psi_1,\rho\rangle$\bork
 $+\langle ({\frac12} S+ {\textstyle{1\over4}}
L_{aa})(\BB \phi-\psi_0),\tilde \BB \rho\rangle
-{1\over2}\Cal{G}_{1,mm}\langle(\Cal{B}\phi-\psi_0),\rho\rangle\}$.
\end{enumerate}\end{theorem}

\section{Remarks} We have presented explicit combinatorial formulas for both the heat content and the heat
trace asymptotics. One of our motivations in computing these invariants was to see if there was a {\bf direct}
combinatorial link between the invariants; there does not seem to be one immediately evident although
techniques used in the computation of both the heat content and the heat trace asymptotics share certain
common features and in principle there are methods which might permit both to be computed simultaneously.
Another example of an asymptotic formula involving geometric data arises from expanding the volume of a tube
of radius $r$ around a submanifold $N$ embedded in an ambient manifold, see for example
\cite{Gray90}. Again, there does not seem to be any direct combinatorial link between these asymptotic
formulae and those we have presented here.

\section{Acknowledgements} It is a pleasant task to thank J.S. Dowker 
for helpful
comments regarding this paper. 
The research of P. Gilkey was partially supported by
the NSF (USA) and the MPI (Leipzig, Germany). 
The research of K. Kirsten was partially supported by the MPI
(Leipzig, Germany) and by EPSRC under Grant No. GR/M45726.
The research of JH. Park was partially
supported by Korea Research 
Foundation Grant (KRF-2000-015-DS0003).
The research of D. Vassilevich was partially supported
by the DFG  project Bo 1112/11-1 (Germany) and by the ESI (Austria).

\end{document}